\documentclass{pasj01}

\def\commenta{$^*$}
\def\commentb{$^\dagger$}
\def\commentc{$^\ddagger$}
\def\commentd{$^\S$}

\newcounter{author}
\def\authorcount#1#2{\refstepcounter{author}\label{#1}
                     \altaffiltext{\ref{#1}}{#2}}

\begin{document}
\SetRunningHead{T. Kato et al.}{PM J03338$+$3320}

\Received{201X/XX/XX}
\Accepted{201X/XX/XX}

\title{PM J03338$+$3320: Long-Period Superhumps in Growing Phase
       Following a Separate Precursor Outburst}

\author{Taichi~\textsc{Kato},\altaffilmark{\ref{affil:Kyoto}*}
        Enrique~de~\textsc{Miguel},\altaffilmark{\ref{affil:Miguel}}$^,$\altaffilmark{\ref{affil:Miguel2}}
        William~\textsc{Stein},\altaffilmark{\ref{affil:Stein}}
        Yutaka~\textsc{Maeda},\altaffilmark{\ref{affil:Mdy}}
        Colin~\textsc{Littlefield},\altaffilmark{\ref{affil:LCO}}
        Seiichiro~\textsc{Kiyota},\altaffilmark{\ref{affil:Kis}}
        Tonny~\textsc{Vanmunster},\altaffilmark{\ref{affil:Vanmunster}}
        Shawn~\textsc{Dvorak},\altaffilmark{\ref{affil:Dvorak}}
        Sergey~Yu.~\textsc{Shugarov},\altaffilmark{\ref{affil:Sternberg}}$^,$\altaffilmark{\ref{affil:Slovak}}
        Eugenia~S.~\textsc{Kalinicheva},\altaffilmark{\ref{affil:Moscow}}
        Roger~D.~\textsc{Pickard},\altaffilmark{\ref{affil:BAAVSS}}$^,$\altaffilmark{\ref{affil:Pickard}}
        Kiyoshi~\textsc{Kasai},\altaffilmark{\ref{affil:Kai}}
        Lewis~M.~\textsc{Cook},\altaffilmark{\ref{affil:LewCook}}
        Hiroshi~\textsc{Itoh},\altaffilmark{\ref{affil:Ioh}}
        Eddy~\textsc{Muyllaert},\altaffilmark{\ref{affil:VVSBelgium}}
}

\authorcount{affil:Kyoto}{
     Department of Astronomy, Kyoto University, Kyoto 606-8502, Japan}
\email{$^*$tkato@kusastro.kyoto-u.ac.jp}

\authorcount{affil:Miguel}{
     Departamento de F\'isica Aplicada, Facultad de Ciencias
     Experimentales, Universidad de Huelva,
     21071 Huelva, Spain}

\authorcount{affil:Miguel2}{
     Center for Backyard Astrophysics, Observatorio del CIECEM,
     Parque Dunar, Matalasca\~nas, 21760 Almonte, Huelva, Spain}

\authorcount{affil:Stein}{
     Center for Backyard Astrophysics, 6025 Calle Paraiso, Las Cruces,
     New Mexico 88012, USA}

\authorcount{affil:Mdy}{
     Kaminishiyamamachi 12-14, Nagasaki, Nagasaki 850-0006, Japan}

\authorcount{affil:LCO}{
     Department of Physics, University of Notre Dame, 
     225 Nieuwland Science Hall, Notre Dame, Indiana 46556, USA}

\authorcount{affil:Kis}{
     Variable Star Observers League in Japan (VSOLJ),
     7-1 Kitahatsutomi, Kamagaya, Chiba 273-0126, Japan}

\authorcount{affil:Vanmunster}{
     Center for Backyard Astrophysics Belgium, Walhostraat 1A,
     B-3401 Landen, Belgium}

\authorcount{affil:Dvorak}{
     Rolling Hills Observatory, 1643 Nightfall Drive,
     Clermont, Florida 34711, USA}

\authorcount{affil:Sternberg}{
     Sternberg Astronomical Institute, Lomonosov Moscow State University, 
     Universitetsky Ave., 13, Moscow 119992, Russia}

\authorcount{affil:Slovak}{
     Astronomical Institute of the Slovak Academy of Sciences,
     05960 Tatranska Lomnica, Slovakia}

\authorcount{affil:Moscow}{
     Faculty of Physics, Lomonosov Moscow State University, 
     Leninskie gory, Moscow 119991, Russia}

\authorcount{affil:BAAVSS}{
     The British Astronomical Association, Variable Star Section (BAA VSS),
     Burlington House, Piccadilly, London, W1J 0DU, UK}

\authorcount{affil:Pickard}{
     3 The Birches, Shobdon, Leominster, Herefordshire, HR6 9NG, UK}

\authorcount{affil:Kai}{
     Baselstrasse 133D, CH-4132 Muttenz, Switzerland}

\authorcount{affil:LewCook}{
     Center for Backyard Astrophysics Concord, 1730 Helix Ct. Concord,
     California 94518, USA}

\authorcount{affil:Ioh}{
     VSOLJ,
     1001-105 Nishiterakata, Hachioji, Tokyo 192-0153, Japan}

\authorcount{affil:VVSBelgium}{
     Vereniging Voor Sterrenkunde (VVS), Moffelstraat 13 3370
     Boutersem, Belgium}



\maketitle

\begin{abstract}
We observed the first-ever recorded outburst of
PM J03338$+$3320, the cataclysmic variable selected
by proper-motion survey.  The outburst was composed of
a precursor and the main superoutburst.  The precursor
outburst occurred at least 5~d before the maximum
of the main superoutburst.  Despite this separation,
long-period superhumps were continuously seen
between the precursor and main superoutburst.
The period of these superhumps is longer than
the orbital period by 6.0(1)\% and can be interpreted
to reflect the dynamical precession rate at 
the 3:1 resonance for a mass ratio of 0.172(4).
These superhumps smoothly evolved into those in
the main superoutburst.  These observations provide
the clearest evidence that the 3:1 resonance is
triggered by the precursor outburst, even if it is
well separated, and the resonance eventually
causes the main superoutburst as predicted by
the thermal-tidal instability model.
The presence of superhumps well before the
superoutburst cannot be explained by alternative
models (the enhanced mass-transfer model and
the pure thermal instability model) and
the present observations give a clear support
to the thermal-tidal instability model.
\end{abstract}

\section{Introduction}

   Cataclysmic variables (CVs) are composed of a white dwarf
and a mass-transferring red (or brown) dwarf filling
the Roche lobe.  The transferred matter forms
an accretion disk.  Dwarf novae are a class of CVs
characterized by outbursts.  SU UMa-type dwarf novae
are a subclass of dwarf novae characterized by
the presence of superhumps and superoutbursts.
Superoutbursts are $\sim$0.5--1.0 mag brighter
than normal outbursts and are accompanied by
superhumps, which have periods (superhump periods: $P_{\rm SH}$)
a few percent longer than the orbital period ($P_{\rm orb}$)
[for general information of CVs, DNe, SU UMa-type
dwarf novae and  superhumps, see e.g.
\citet{war95book}; \citet{hel01book}].

   Although there had been a long debate regarding
the origin of superoutbursts and superhumps,
it is now widely believed that superhumps are
a result of the eccentric (or flexing) deformation
of the accretion disk caused by the 3:1 resonance
(\cite{whi88tidal}; \cite{hir90SHexcess};
\cite{lub91SHa}; \cite{lub91SHb}).
\citet{osa89suuma} proposed a model [thermal-tidal
instability (TTI) model] to explain the occurrence
of a superoutburst after a sequence of normal
outbursts when the disk radius reaches the 3:1
resonance.  \citet{osa03DNoutburst} refined
this TTI model to explain various types of
superoutbursts.  In some systems [usually with
lower mass-transfer rates ($\dot{M}$)], the mass
stored in the disk is sufficient to trigger
a superoutburst without experiencing normal
outbursts.  The extreme cases are WZ Sge-type
dwarf novae (see \citet{kat15wzsge} for a modern
review).  This TTI model predicted the systematic
variation of the disk radius over successive
outbursts and superoutbursts.  This prediction was
finally confirmed in Kepler observations, which
led to the strongest ever support to the TTI model
\citep{osa13v1504cygKepler}.

   On the other hand, it has been demonstrated
that periods of superhumps systematically vary
\citep{Pdot}: stage A superhumps (long, constant superhump period),
stage B superhumps (short superhump period with systematic period
variations) and stage C superhumps (constant period shorter than
those of stage B superhumps typically by 0.5\%).
The superhump period reflects the precession
rate of the eccentric disk, which is mainly
a combination of prograde dynamical precession by
the gravitational field of the secondary
and the retrograde precession by the pressure gradient
in the disk (e.g. \cite{lub92SH}).
It has been proposed that stage A superhumps
corresponds to the growing phase of the 3:1 resonance
(\cite{osa13v344lyrv1504cyg}; \cite{kat13qfromstageA})
and the superhump period in this stage reflects
the dynamical precession rate at the 3:1 resonance.
This expectation was successfully used to determine
mass-ratios ($q$) in many systems \citep{kat13qfromstageA}.

   In the standard TTI model, superoutbursts are
triggered by a normal outburst during which the disk radius
reaches that of the 3:1 resonance.
In actual SU UMa-type dwarf novae, there is diversity
how superoutbursts start (cf. \cite{mar79superoutburst}).
In some cases, there are superoutbursts preceded by
widely separated precursor outbursts.
Although the TTI model predicts that the 3:1 resonance
(and resultant superhumps) starts to grow during
the preceding precursor or sometime after
the precursor, observational clues for this interpretation
have been limited since it is difficult to make
high-quality time-resolved photometry {\it before}
superoutbursts start.  The only exception has been continuous
observations by the Kepler satellite (\cite{bor10Keplerfirst};
\cite{Kepler}).  Kepler recorded a deep dip between
the precursor outburst and main superoutburst in V1504 Cyg.
Although a frequency analysis by \citet{osa14v1504cygv344lyrpaper3}
strongly favored the interpretation by the TTI model,
the case of V1504 Cyg was not ideal in that
the precursor outburst was not sufficiently separated
and that the superhump signal was not continuously
present.  Here we report an ideal case to test
this interpretation: the 2015 superoutburst of PM J03338$+$3320.

\section{PM J03338$+$3320}

   PM J03338$+$3320 is a CV selected by its high
proper motion \citep{ski14SuperblinkCVs}.\footnote{
   Although the name PM I03338$+$3320 was used in
   \citet{ski14SuperblinkCVs}, we use the name used
   in SIMBAD, conforming the nomenclature convention of
   the International Astronomical Union.
   The acronym LSPM was also used in \citet{ski11j0333}
   and the name LSPM J03338$+$3320 was used in
   VSNET reports.
}
\citet{ski14SuperblinkCVs} indicated that PM J03338$+$3320
has doubly peaked emission lines of hydrogen and He\textsc{I}
superimposed on a blue continuum.  The spectrum
was typical for a dwarf nova in quiescence.
\citet{ski14SuperblinkCVs} also obtained an orbital
period of 0.06663(7)~d by a radial-velocity study.
The period was suggestive of an SU UMa-type dwarf nova.

   On 2015 November 28, E. Muyllaert detected
the first-ever outburst at an unfiltered CCD
magnitude of 14.58 (cvnet-outburst 6781).
Although subsequent observations detected
superhump-like modulations (vsnet-alert 19303),
the object rapidly faded.  Although there was
some suspicion of a normal outburst,
the long period compared to the known orbital period
strongly suggested that these modulations were
superhumps.  The object stayed around 17 mag
for three nights and it continuously showed
these long-period superhumps (vsnet-alert 19310).
The object eventually entered the main superoutburst
on December 2--3.

\section{Observation and Analysis}\label{sec:obs}

   The data were obtained under campaigns led by 
the VSNET Collaboration \citep{VSNET}.
We also used the public data from 
the AAVSO International Database\footnote{
   $<$http://www.aavso.org/data-download$>$.
}.

Time-resolved observations were performed in 15 different
locations by using 30cm-class telescopes (supplementary table 1).
The data analysis was performed just in the same way described
in \citet{Pdot} and \citet{Pdot6} and we mainly used
R software\footnote{
   The R Foundation for Statistical Computing:\\
   $<$http://cran.r-project.org/$>$.
} for data analysis.
In de-trending the data, we divided the data into
four segments in relation to the outburst phase and
used locally-weighted polynomial regression 
(LOWESS: \cite{LOWESS}).
The times of superhumps maxima were determined by
the template fitting method as described in \citet{Pdot}.
The times of all observations are expressed in 
barycentric Julian Days (BJD).

   We used phase dispersion minimization (PDM; \cite{PDM})
for period analysis and 1$\sigma$ errors for the PDM analysis
was estimated by the methods of \citet{fer89error} and \citet{Pdot2}.

\section{Results}\label{sec:results}

\subsection{Outburst Light Curve}\label{sec:outburst}

\begin{figure*}
  \begin{center}
    \FigureFile(110mm,170mm){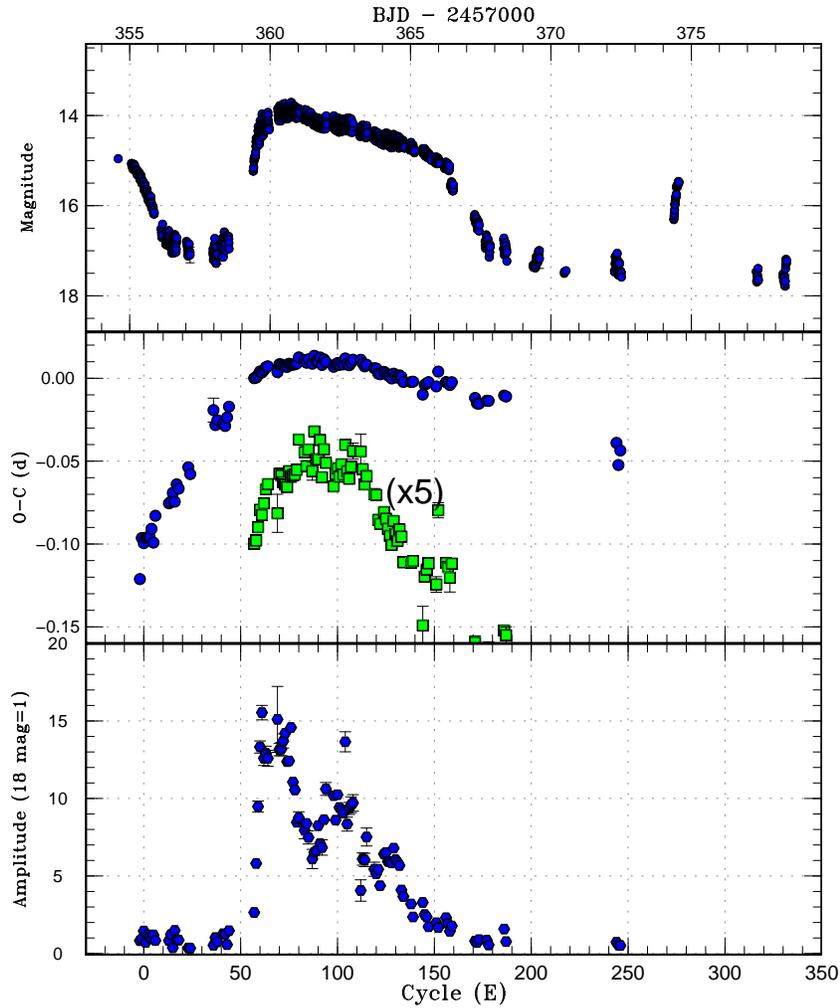}
  \end{center}
  \caption{$O-C$ diagram of superhumps in PM J03338$+$3320 (2015).
     (Upper:) Light curve.  The data were binned to 0.0069~d.
     There was at least one post-superoutburst rebrightening
     on BJD 2457374--2457375.
     (Middle:) $O-C$ diagram (filled circles).
     We used a period of 0.0690~d for calculating the $O-C$ residuals.
     Filled squares are enlarged by five times in the $O-C$
     values and shifted arbitrarily to better visualize
     the stage transitions.
     (Lower:) Amplitudes of superhumps.  The scale is linear
     and the pulsed flux is shown in a unit corresponding
     to 18 mag = 1.
  }
  \label{fig:j0333humpall}
\end{figure*}

   As shown in the upper panel of
figure \ref{fig:j0333humpall}, this
object showed a separate precursor outburst which
occurred at least 5~d before the peak of the main
superoutburst.  The main superoutburst lasted for
$\sim$7~d.  Between the precursor and main
superoutburst, the object stayed at $\sim$16.8,
$\sim$0.5 mag brighter than in quiescence.
There was at least one post-superoutburst rebrightening
on BJD 2457374--2457375, 7~d after the rapid fading.
The object was also caught during the fading part
of another outburst on BJD 2457400 (E. de Miguel).
It was not certain whether this outburst was the second
rebrightening or the first normal outburst of the next
supercycle.

\subsection{Superhumps}\label{sec:superhumps}

\begin{figure*}
  \begin{center}
    \FigureFile(140mm,100mm){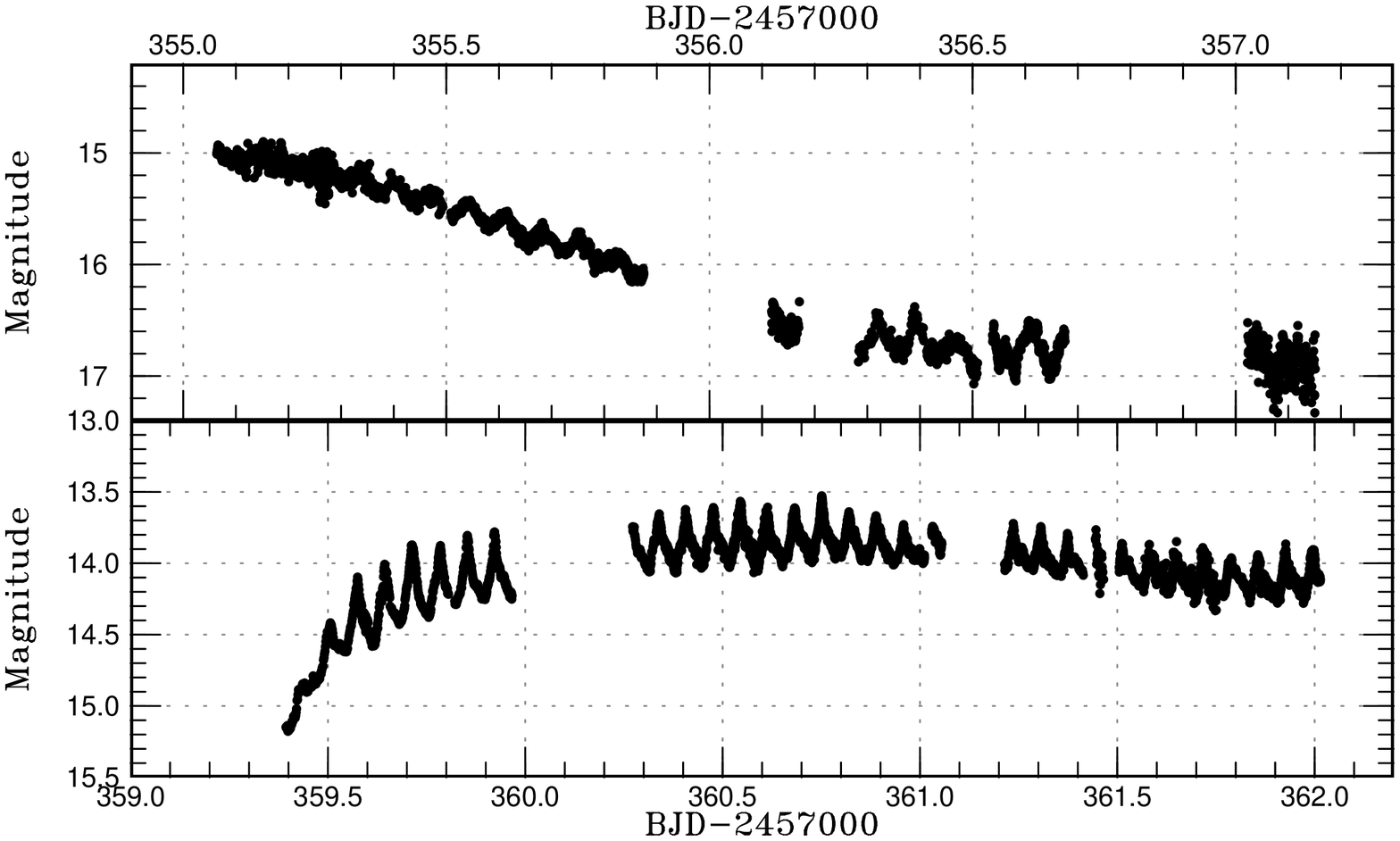}
  \end{center}
  \caption{Superhumps in PM J03338$+$3320 (2015).
  (Upper) Fading part of the precursor.
  (Lower) Initial part of the main superoutburst.
  }
  \label{fig:j0333humpgrow}
\end{figure*}

\begin{figure}
  \begin{center}
    \FigureFile(88mm,110mm){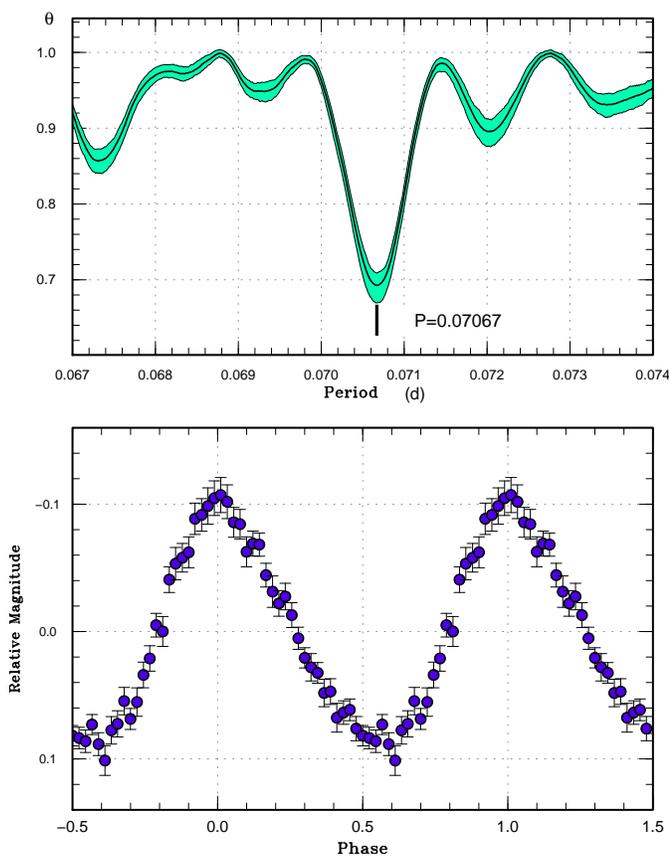}
  \end{center}
  \caption{Superhumps in PM J03338$+$3320 (2015) from
     the precursor to the main superoutburst.
     The data before BJD 2457360 were used.
     (Upper): PDM analysis.
     (Lower): Phase-averaged profile.}
  \label{fig:j0333shapdm}
\end{figure}

\begin{figure}
  \begin{center}
    \FigureFile(88mm,110mm){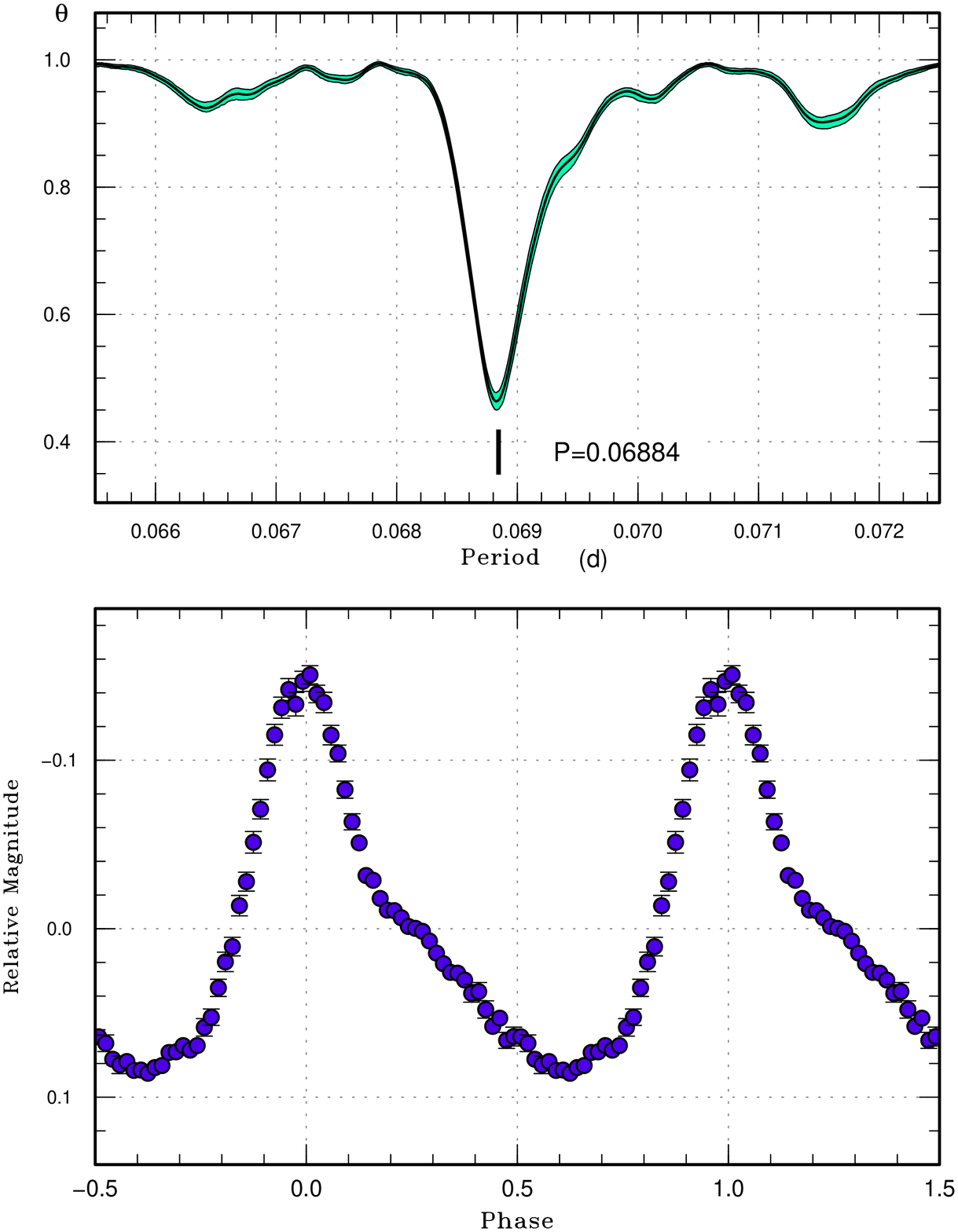}
  \end{center}
  \caption{Superhumps in PM J03338$+$3320 (2015)
     during the main superoutburst after transition
     tp stage B superhumps.
     The data after BJD 2457360 were used.
     (Upper): PDM analysis.
     (Lower): Phase-averaged profile.}
  \label{fig:j0333shbpdm}
\end{figure}

   Superhumps were continuously detected already
during the fading branch of the precursor outburst
(figure \ref{fig:j0333humpgrow}).
The superhumps before the rise to the main
superoutburst had a long period [0.07066(1)~d
in average; figure \ref{fig:j0333shapdm};
supplementary table 2].
As shown in the $O-C$ diagram in
figure \ref{fig:j0333humpall}, these superhumps
showed no period variation.  These superhumps
smoothly evolved into superhumps during
the main superoutburst.  The period became slightly
shorter when the object approached the maximum of
the superoutburst (around $E$=60--65 in
figure \ref{fig:j0333humpall}).  Superhumps following
this phase had a shorter, relatively constant
period of 0.06902(2)~d between $E$=70 and $E$=115
in figure \ref{fig:j0333humpall}.
The period then shortened to another constant one
at 0.06876(3)~d (figure \ref{fig:j0333shbpdm},
mean profile after $E$=70; supplementary table 3).
These superhumps persisted at least before
the rebrightening.
We identified these three stages
as stage A, B and C as introduced in \citet{Pdot}.
The most striking feature is the continuous presence
of stage A superhumps between the separate precursor
and the main superoutburst.
The apparent lack of a phase jump in stage C
superhumps when the superoutburst terminated
is also worth mentioning.

   There is likely modulations of the superhump
amplitudes with a period of $\sim$30 cycles
(see lower panel of figure \ref{fig:j0333humpall}).
This is most likely a beat phenomenon between
the orbital period and superhump period
(the expected beat period is 2.15~d = 31 cycles).
The presence of the beat phenomenon is also
consistent with the likely high inclination
inferred from doubly peaked emission lines in spectroscopy.

\section{Discussion}\label{sec:discussion}

\subsection{Separate precursor and stage A superhumps}\label{sec:stagea}

   It has been known that there is a continuous
sequence of precursor-main superoutburst in VW Hyi
\citep{mar79superoutburst}.
As discussed in subsection 2.2 in
\citet{osa14v1504cygv344lyrpaper3}, this phenomenon
can be understood in the framework of the TTI model
that the 3:1 resonance is reached during the precursor
outburst and superhumps start to grow, finally triggering
the main superoutburst.  In the case of Kepler data
in V1504 Cyg, a frequency analysis has shown that
long-period growing superhumps (stage A superhumps)
were indeed present between the precursor
outburst and main superoutburst
\citep{osa14v1504cygv344lyrpaper3}.
In the case of V1504 Cyg, however, the superhump
waves was rather irregular in shape and the signal almost
disappeared during the rising branch to the main
superoutburst [see figure 5
in \citet{osa14v1504cygv344lyrpaper3}].
This made it impossible to make an $O-C$ analysis
to see whether the superhumps between the precursor
outburst and main superoutburst have the continuous
properties as those recorded during the main
superoutburst.

   The present case of PM J03338$+$3320 provides
an ideal opportunity in studying the properties of
the superhumps between the precursor outburst and
main superoutburst: the superhump signals were strong
and regular and we could measure individual maxima.
The result (middle panel of figure \ref{fig:j0333humpall})
clearly indicates that superhumps had the same period
and the phase since the precursor until the peak
of the main superoutburst.  We can now safely say
that stage A superhumps were persistently present
following the precursor outburst and they smoothly
evolved into stage A superhumps during the main
superoutburst.  The present case is more extreme
than in V1504 Cyg in that the precursor is more
isolated from the main superoutburst.
This finding gives a strong support to the TTI model.

   The striking finding is the constancy of the period
during and after the precursor outburst despite
that the system brightness greatly decreased.
As discussed in \citet{kat13qfromstageA},
the fractional superhump excess in frequency
$\epsilon^* = 1-P_{\rm orb}/P_{\rm SH}$
has a functional form of $\epsilon^* = Q(q) R(r)$,
where $r$ is the disk radius,
when the pressure effect can be neglected.
When the system is faint (as in the faint state
between the precursor outburst and main superoutburst),
we can safely neglect the pressure effect
and the $\epsilon^*$ can be used to estimate
the change of the radius (cf. \cite{osa14v1504cygv344lyrpaper3}).
The measured $\epsilon^*$ for superhumps between
the precursor outburst and main superoutburst
was large and constant and the disk radius needs
to be constant.  We consider that this radius
represents the radius of the eccentric part of
the disk.  As shown in subsection \ref{sec:massratio},
this radius is compatible with that of the 3:1 resonance
assuming a reasonable $q$ for the orbital period.
We now have evidence
that the eccentric disk at the 3:1 resonance is
continuously present between the precursor outburst and
main superoutburst, which was supposed, but not
sufficiently verified yet, in the TTI model.

\subsection{Mass ratio from stage A superhumps}\label{sec:massratio}

   \citet{kat13qfromstageA} proposed a method to determine
$q$ from $\epsilon^*$ for stage A superhumps.
In the present case,
the observed $\epsilon^*$ = 0.0604(13) corresponds to 
$q$=0.172(4).  This value is somewhat large for
a $P_{\rm orb}$=0.06663~d object (cf. figure 4
in \citet{kat13qfromstageA}).  This large estimated $q$
safely excludes a possibility that the radius where
stage A superhumps arose was somewhere inside
the 3:1 resonance --- if the radius is smaller,
it requires an even unlikely larger $q$
for this $P_{\rm orb}$.

\subsection{Difference from V1504 Cyg}\label{sec:v1504dif}

   In the case of Kepler data of V1504 Cyg,
although frequency analysis detected stage A
superhumps between the precursor and main superoutburst,
individual hump profiles were rather complex and
we could not sufficiently measure individual maxima.
The profiles in PM J03338$+$3320, however, were
much more clearer and we could measure times of
all observed superhumps in the same outburst phase.
The difference may be due to the difference
in the mass-transfer rate.  In the case of V1504 Cyg,
the frequent normal outbursts and short supercycle
(cf. \cite{osa13v1504cygKepler}; \cite{osa13v344lyrv1504cyg};
\cite{osa14v1504cygv344lyrpaper3}) suggests a high
mass-transfer rate.  In contrast, PM J03338$+$3320
apparently has much infrequent outbursts.
Indeed, no outbursts were recorded in observations on
90 different nights spanning for BJD 2453642--2456593
in the CRTS data \citep{CRTS}.
In the case of V1504 Cyg, the strong mass-accretion
flow may have masked low-amplitude stage A superhumps
before the main superoutburst by resulting
strong flickering.
Although the stage A superhump method is expected to
work to determine the mass ratio when superhumps
start to evolve before the main superoutburst
(as in the present case), the best application may be
sought for low-$\dot{M}$ objects.  This interpretation
needs to be confirmed by further observations
in different objects.

\subsection{Growth time of stage A superhumps}\label{sec:stageagrowth}

   \citet{kat15wzsge} suggested that the growth time
of stage A superhumps in WZ Sge-type dwarf novae
can be a good measure of the growth of
the 3:1 resonance, which is theoretically expected
to be proportional to $q^2$ (\cite{lub91SHa}; \cite{lub91SHb}).
In WZ Sge-type dwarf novae, $q$=0.06 roughly corresponds
to the growth time of 60 cycles and the time is
usually much shorter in ordinary SU UMa-type
dwarf novae.  \citet{kat16v1006cyg} claimed that
in high $q$ systems close to the stability limit
of the 3:1 resonance also show slow growth of
superhumps, requiring time comparable to
WZ Sge-type dwarf novae.
The present case of PM J03338$+$3320 required
at least 60 cycles, which is comparable to
low-$q$ WZ Sge-type dwarf novae.  Since PM J03338$+$3320
does not have a high $q$ close to the stability limit,
this long growth time requires another explanation.
The main difference of PM J03338$+$3320
from WZ Sge-type dwarf novae and other SU UMa-type
dwarf novae is the widely separated precursor outburst.  
Stage A superhumps showed little tendency to grow
in amplitude during this phase (lower panel of
figure \ref{fig:j0333humpall}), and it is likely
that stage A superhumps grow slowly near the quiescent
state.  Since the growth of stage A superhumps,
and finally the spread into the entire disk
would require viscous spread of the eccentric region,
it may be that the low-viscosity state near quiescence
requires more time than in the outbursting disk
as in WZ Sge-type dwarf novae and SU UMa-type
dwarf novae without separate precursor outbursts.
It means that the duration of stage A would not be
a good measure of $q$ in systems with separate
precursor outbursts.  This would, in turn, explain
the diversity in the intervals between
the precursor outburst and the main superoutburst
(cf. ranging from almost continuous transition
to intervals as long as 10~d in
QZ Vir (1998, \cite{ohs11qzvir}) and
11~d in V699 Oph (2001, \cite{Pdot}, see
\cite{osa14v1504cygv344lyrpaper3})
since the growth time of stage A superhumps
is not a unique function of $q$ near quiescent state.

\subsection{Cases of enhanced mass-transfer and pure thermal instability models}\label{sec:pureTI}

   Up to this subsection, we considered the TTI model
to interpret the observations.  As reviewed in
\citet{osa13v1504cygKepler}, there are currently
three models to explain superoutbursts and supercycles.
In addition to the TTI model, there are the enhanced
mass-transfer (EMT) model advocated by Smak
(\cite{sma91suumamodel}, \cite{sma04EMT}, \cite{sma08zcha}), 
and the pure thermal instability model by Cannizzo
in the original form (\cite{can10v344lyr}, \cite{can12v344lyr}).

   In the present case, superhumps appeared well before
the main superoutburst, which excludes the EMT model
as already discussed in \citet{osa13v1504cygKepler}.
We consider here if the present observations can be
explained by the pure thermal instability model.
As reviewed in \citet{osa13v1504cygKepler},
this model claims that normal outbursts and superoutbursts
are equivalent to ``narrow'' and ``wide'' outbursts seen in
SS Cyg-type dwarf novae and that superhumps are excited
as the result of the expansion of the disk during
wide outbursts.  In this model, heating wave from
the inner part of the disk is reflected before reaching
the outer edge of the disk in normal outbursts
[see figures 3, 4 in \citet{can10v344lyr};
this type of outburst corresponds to type `Bb'
in the classification by \citet{sma84DI}].
Only in superoutbursts the heating wave reaches the outer edge.
In the pure thermal instability model, precursor outbursts
(or shoulders) are formed since the speed of the heating wave
becomes slower when it passes through the disk mass
which has been accumulated during repeated cycles
of normal outbursts preceding the superoutburst.
During such precursors, the heating wave should not
be reflected since such reflection will quench
the outburst (see subsection 3.2 in \cite{osa13v1504cygKepler}).
There should not be a deep dip between the precursor
(shoulder) and the peak of the main superoutburst
(see also subsection 2.2 in \cite{osa14v1504cygv344lyrpaper3}).
Its natural consequence is that it is
impossible to reproduce a separate precursor as
seen in the present observations.
Another difficulty in the pure thermal instability model
is that the disk does not expand during normal
outbursts or the precursor (shoulder) phase,
since the heating wave does not reach the outer edge
of the disk.  In this model, the disk cannot reach
the 3:1 resonance before the heating wave reaches
close to the outer edge, i.e. around the peak
of the superoutburst.  The presence of long-period superhumps
well before the superoutburst, which is
the direct consequence of the disk reaching
the 3:1 resonance, cannot be explained by
the pure thermal instability model and the TTI model
is the only model which can explain the present
observations.

   In this subsection, we used ``in the original form''
for the pure thermal instability model by Cannizzo.
This was because \citet{can15v1504cygv344lyrproc}
recently introduced a pure thermal instability model allowing
the radius variation of the disk and claimed that
this model could reproduce the SU UMa-type supercycle.
Although this model appears to have successfully
reproduced the presence of a precursor outburst,
the reason why such a precursor was made possible
is unclear since this paper is a conference proceeding and
no details of the model were given.
It would be better for a full article on this model
to appear to make a fair comparison between the TTI model
with this new extension of the pure thermal instability model
and we restrict our discussion to the original
pure thermal instability model.
We can, however, point out that the model in
\citet{can15v1504cygv344lyrproc} would
predict the constant disk radius during the superoutburst
(as in figure 2 in \cite{can15v1504cygv344lyrproc}),
which contradicts the observation, such as the decrease
in the disk radius measured in eclipsing systems
(see e.g. figure 10 in \cite{osa14v1504cygv344lyrpaper3})
and the systematic decrease of the superhump periods
(i.e. precession rates) between superhump stages B and C.

\section{Summary}\label{sec:summary}

   We observed the 2015 superoutburst of PM J03338$+$3320.
The superoutburst was preceded by a separate
precursor outburst which occurred at least 5~d
before the maximum of the main superoutburst.
Superhumps were continuously present during
the fading branch of the precursor and persisted
until the rise to the main superoutburst.
The $O-C$ analysis has shown that these superhumps
(stage A superhumps)
have continuous phases and a constant period
all the time before the maximum of the main superoutburst.
The period was very long, 0.07066(1)~d, 6.0(1)\%
longer than the orbital period, and can be interpreted
to reflect the dynamical precession rate at
the 3:1 resonance for a mass ratio of 0.172(4).
This result provides the clearest evidence that 
the 3:1 resonance starts to operate around 
the precursor outburst, even if it is well separated,
and this resonance eventually
results the main superoutburst as predicted by
the thermal-tidal instability model.
These superhumps took long time (more than 60 cycles)
to evolve, suggesting that stage A superhumps
persist longer time (or take longer time to
fully evolve) when the system is near
quiescence than in the outburst state.
This could explain the wide variety of intervals between
the precursor and the main superoutburst.
The presence of superhumps well before the
superoutburst cannot be explained by alternative
models (the enhanced mass-transfer model and
the pure thermal instability model) and
the present observations give a clear support
to the thermal-tidal instability model.

\section*{Acknowledgments}

This work was supported by the Grant-in-Aid
``Initiative for High-Dimensional Data-Driven Science through Deepening
of Sparse Modeling'' (25120007) from the Ministry of Education,
Culture, Sports, Science and Technology (MEXT) of Japan.
We acknowledge with thanks the variable star
observations from the AAVSO International Database contributed by
observers worldwide and used in this research. 
The authors are grateful to observers of VSNET Collaboration and
VSOLJ observers who supplied vital data.
We are grateful to the anonymous referee for drawing
our attention to the pure thermal instability model
and Prof. Yoji Osaki for helpful comments.
This work also was partially supported by grants
of RFBR 15-32-50920 and 15-02-06178,
14-02-00825 and by the VEGA grant No. 2/0002/13.

\section*{Supporting information}

Additional supporting information in the online version
of this article in PASJ: Tables 1, 2, 3.

\setcounter{figure}{1}

\setcounter{table}{0}

\begin{table*}
\caption{Log of observations of PM J03338$+$3320 (2015)}
\begin{center}
\begin{tabular}{cccccccccc}
\hline
Start\commenta & End\commenta & $N$\commentb & Code\commentc & filter\commentd &
Start\commenta & End\commenta & $N$\commentb & Code\commentc & filter\commentd \\
\hline
57355.0583 & 57355.2732 &  385 &  Mdy &  C & 57363.1758 & 57363.2750 &   54 &  Shu &  C \\
57355.1568 & 57355.3496 &  215 &  Ioh &  C & 57363.2322 & 57363.3427 &  149 &  RAE &  V \\
57355.2755 & 57355.6777 &  482 &  deM &  C & 57363.2411 & 57363.4529 &  281 &  CDZ &  C \\
57355.5031 & 57355.8485 &  390 &  BJA &  C & 57363.6883 & 57363.9489 &  344 &  SWI &  V \\
57355.7491 & 57355.8698 &  295 &  COO &  C & 57364.0347 & 57364.3080 &  684 &  Mdy &  C \\
57356.1126 & 57356.1654 &  120 &  Mdy &  C & 57364.3162 & 57364.4267 &  113 &  Van &  C \\
57356.2777 & 57356.5040 &  221 &  deM &  C & 57364.4108 & 57364.5892 &  152 &  Shu &  C \\
57356.5324 & 57356.6701 &  376 &  LCO &  C & 57364.5434 & 57364.7917 &  304 &  DKS &  C \\
57357.0169 & 57357.1456 &  262 &  Mdy &  C & 57364.9669 & 57365.1251 &  210 &  Mdy &  C \\
57357.9564 & 57358.1116 &  305 &  Mdy &  C & 57365.0079 & 57365.1449 &  301 &  Kis &  C \\
57358.2716 & 57358.5370 &  147 &  deM &  C & 57365.4313 & 57365.6558 &  310 &  RPc &  V \\
57359.3884 & 57359.5962 &  166 &  deM &  C & 57365.5197 & 57365.7292 &  252 &  DKS &  C \\
57359.5608 & 57359.9611 &  500 &  SWI &  V & 57365.8744 & 57366.0356 &  374 &  Kis &  C \\
57360.2665 & 57360.6119 &  401 &  deM &  C & 57366.2104 & 57366.3845 &  220 &  Van &  C \\
57360.3357 & 57360.6309 &  244 &  Shu &  C & 57366.4192 & 57366.5233 &  134 &  Kai &  C \\
57360.4993 & 57360.9179 &  920 &  BJA &  C & 57367.2630 & 57367.4397 &  210 &  deM &  C \\
57360.5564 & 57360.9585 &  520 &  SWI &  V & 57367.6499 & 57367.8332 &  224 &  DKS &  C \\
57360.5627 & 57360.7461 &  459 &  LCO &  C & 57368.3059 & 57368.4264 &  129 &  RPc &  C \\
57360.7774 & 57360.9686 &  104 &  COO &  C & 57369.3588 & 57369.5860 &   41 &  Shu &  C \\
57360.9562 & 57361.0502 &  178 &  Kis &  C & 57370.4704 & 57370.4794 &    4 &  Shu &  C \\
57361.2089 & 57361.3714 &  254 &  Shu &  C & 57372.2571 & 57372.5108 &  210 &  deM &  C \\
57361.2356 & 57361.4084 &  220 &  Van &  C & 57374.3709 & 57374.5600 &  145 &  deM &  C \\
57361.3066 & 57361.4600 &   40 &  CDZ &  C & 57377.3073 & 57377.3875 &   65 &  deM &  C \\
57361.4993 & 57361.7734 &  600 &  BJA &  C & 57378.2637 & 57378.3963 &  111 &  deM &  C \\
57361.5617 & 57361.7452 &  243 &  CDZ &  C & 57382.2677 & 57382.2968 &   25 &  deM &  C \\
57361.7101 & 57361.9306 &  292 &  SWI &  V & 57386.3241 & 57386.3386 &   13 &  deM &  C \\
57361.8531 & 57362.0093 &  363 &  Kis &  C & 57393.2677 & 57393.3415 &   60 &  deM &  C \\
57362.2419 & 57362.6629 &  557 &  CDZ &  C & 57400.2860 & 57400.2922 &    5 &  deM &  C \\
57362.3846 & 57362.6203 &  236 &  Van &  C & 57401.2771 & 57401.2877 &    8 &  deM &  C \\
57362.4499 & 57362.6643 &  287 &  Kai &  V & 57403.2826 & 57403.2872 &    4 &  deM &  C \\
57362.4824 & 57362.6720 &  753 &  LCO &  C & -- & -- & -- & -- & -- \\
\hline
  \multicolumn{10}{l}{\commenta JD$-$2400000.} \\
  \multicolumn{10}{l}{\commentb Number of observations.} \\
  \multicolumn{10}{l}{\parbox{400pt}{\commentc Key to observers:
COO (L. Cook),
DKS (S. Dvorak),
deM (E. de Miguel),
Ioh (H. Itoh),
Kai (K. Kasai),
Kis (S. Kiyota),
LCO (C. Littlefield),
Mdy (Y. Maeda),
RPc (R. Pickard),
Shu (S. Shugarov team),
SWI (W. Stein),
Van (T. Vanmunster),
BJA, CDZ, RAE (AAVSO observations).
  }} \\
  \multicolumn{10}{l}{\commentd The filter name C represents unfiltered observations.} \\
\end{tabular}
\end{center}
\end{table*}

\clearpage

\begin{table*}
\caption{Superhump maxima of PM J03338$+$3320 (2015) before the main superoutburst}
\begin{center}
\begin{tabular}{rp{55pt}p{40pt}r@{.}lr}
\hline
\multicolumn{1}{c}{$E$} & \multicolumn{1}{c}{max\commenta} & \multicolumn{1}{c}{error} & \multicolumn{2}{c}{$O-C$\commentb} & \multicolumn{1}{c}{$N$\commentc} \\
\hline
0 & 57355.2416 & 0.0028 & $-$0&0164 & 152 \\
1 & 57355.3354 & 0.0014 & 0&0064 & 126 \\
2 & 57355.4013 & 0.0009 & 0&0014 & 63 \\
3 & 57355.4735 & 0.0019 & 0&0027 & 59 \\
4 & 57355.5429 & 0.0003 & 0&0012 & 147 \\
5 & 57355.6120 & 0.0004 & $-$0&0006 & 147 \\
6 & 57355.6860 & 0.0006 & 0&0025 & 106 \\
7 & 57355.7467 & 0.0013 & $-$0&0078 & 116 \\
8 & 57355.8319 & 0.0011 & 0&0065 & 186 \\
15 & 57356.3223 & 0.0007 & 0&0006 & 52 \\
16 & 57356.3923 & 0.0005 & $-$0&0003 & 55 \\
17 & 57356.4667 & 0.0014 & 0&0031 & 57 \\
18 & 57356.5304 & 0.0019 & $-$0&0041 & 66 \\
19 & 57356.6098 & 0.0005 & 0&0044 & 154 \\
20 & 57356.6762 & 0.0009 & $-$0&0001 & 77 \\
25 & 57357.0341 & 0.0033 & 0&0033 & 80 \\
26 & 57357.0990 & 0.0030 & $-$0&0027 & 116 \\
38 & 57357.9656 & 0.0072 & 0&0129 & 45 \\
39 & 57358.0256 & 0.0010 & 0&0020 & 125 \\
40 & 57358.0975 & 0.0014 & 0&0030 & 99 \\
43 & 57358.3023 & 0.0008 & $-$0&0049 & 42 \\
44 & 57358.3700 & 0.0011 & $-$0&0081 & 31 \\
45 & 57358.4442 & 0.0024 & $-$0&0049 & 36 \\
46 & 57358.5197 & 0.0027 & $-$0&0003 & 22 \\
\hline
  \multicolumn{6}{l}{\commenta BJD$-$2400000.} \\
  \multicolumn{6}{l}{\commentb Against max $= 2457355.2581 + 0.070910 E$.} \\
  \multicolumn{6}{l}{\commentc Number of points used to determine the maximum.} \\
\end{tabular}
\end{center}
\end{table*}

\clearpage

\begin{table*}
\caption{Superhump maxima of PM J03338$+$3320 (2015) during the main superoutburst and post-superoutburst}
\begin{center}
\begin{tabular}{rp{55pt}p{40pt}r@{.}lrrp{55pt}p{40pt}r@{.}lr}
\hline
\multicolumn{1}{c}{$E$} & \multicolumn{1}{c}{max\commenta} & \multicolumn{1}{c}{error} & \multicolumn{2}{c}{$O-C$\commentb} & \multicolumn{1}{c}{$N$\commentc} & \multicolumn{1}{c}{$E$} & \multicolumn{1}{c}{max\commenta} & \multicolumn{1}{c}{error} & \multicolumn{2}{c}{$O-C$\commentb} & \multicolumn{1}{c}{$N$\commentc} \\
\hline
0 & 57359.4339 & 0.0006 & $-$0&0152 & 45 & 55 & 57363.2400 & 0.0021 & 0&0085 & 85 \\
1 & 57359.5032 & 0.0004 & $-$0&0146 & 42 & 56 & 57363.3068 & 0.0006 & 0&0066 & 150 \\
2 & 57359.5738 & 0.0003 & $-$0&0127 & 91 & 57 & 57363.3740 & 0.0007 & 0&0050 & 75 \\
3 & 57359.6449 & 0.0003 & $-$0&0104 & 65 & 58 & 57363.4440 & 0.0007 & 0&0062 & 56 \\
4 & 57359.7133 & 0.0003 & $-$0&0108 & 72 & 62 & 57363.7178 & 0.0007 & 0&0050 & 69 \\
5 & 57359.7837 & 0.0003 & $-$0&0092 & 66 & 63 & 57363.7867 & 0.0004 & 0&0051 & 72 \\
6 & 57359.8544 & 0.0003 & $-$0&0072 & 71 & 64 & 57363.8527 & 0.0005 & 0&0023 & 73 \\
7 & 57359.9240 & 0.0004 & $-$0&0064 & 73 & 65 & 57363.9212 & 0.0004 & 0&0021 & 73 \\
12 & 57360.2655 & 0.0023 & $-$0&0088 & 25 & 67 & 57364.0607 & 0.0004 & 0&0039 & 116 \\
13 & 57360.3393 & 0.0003 & $-$0&0038 & 85 & 68 & 57364.1289 & 0.0003 & 0&0034 & 138 \\
14 & 57360.4081 & 0.0003 & $-$0&0038 & 117 & 69 & 57364.1966 & 0.0003 & 0&0024 & 143 \\
15 & 57360.4763 & 0.0003 & $-$0&0043 & 100 & 70 & 57364.2648 & 0.0003 & 0&0018 & 140 \\
16 & 57360.5450 & 0.0002 & $-$0&0044 & 266 & 71 & 57364.3327 & 0.0005 & 0&0009 & 67 \\
17 & 57360.6137 & 0.0002 & $-$0&0044 & 418 & 72 & 57364.4046 & 0.0005 & 0&0040 & 64 \\
18 & 57360.6846 & 0.0002 & $-$0&0023 & 341 & 73 & 57364.4720 & 0.0004 & 0&0027 & 46 \\
19 & 57360.7529 & 0.0002 & $-$0&0028 & 216 & 74 & 57364.5402 & 0.0004 & 0&0020 & 71 \\
20 & 57360.8221 & 0.0002 & $-$0&0024 & 224 & 75 & 57364.6105 & 0.0004 & 0&0037 & 79 \\
21 & 57360.8912 & 0.0002 & $-$0&0021 & 223 & 76 & 57364.6787 & 0.0005 & 0&0030 & 67 \\
22 & 57360.9608 & 0.0003 & $-$0&0012 & 122 & 77 & 57364.7446 & 0.0006 & 0&0002 & 66 \\
23 & 57361.0334 & 0.0007 & 0&0026 & 73 & 81 & 57365.0205 & 0.0006 & 0&0010 & 150 \\
26 & 57361.2389 & 0.0005 & 0&0018 & 117 & 82 & 57365.0898 & 0.0005 & 0&0015 & 205 \\
27 & 57361.3062 & 0.0003 & 0&0004 & 180 & 87 & 57365.4270 & 0.0023 & $-$0&0052 & 38 \\
28 & 57361.3772 & 0.0005 & 0&0026 & 102 & 88 & 57365.5019 & 0.0006 & 0&0010 & 81 \\
30 & 57361.5126 & 0.0011 & 0&0004 & 78 & 89 & 57365.5717 & 0.0005 & 0&0020 & 145 \\
31 & 57361.5864 & 0.0004 & 0&0054 & 184 & 90 & 57365.6415 & 0.0006 & 0&0030 & 133 \\
32 & 57361.6520 & 0.0004 & 0&0023 & 193 & 94 & 57365.9149 & 0.0009 & 0&0014 & 127 \\
33 & 57361.7210 & 0.0004 & 0&0025 & 235 & 95 & 57365.9929 & 0.0009 & 0&0106 & 128 \\
34 & 57361.7923 & 0.0004 & 0&0051 & 104 & 99 & 57366.2625 & 0.0006 & 0&0051 & 69 \\
35 & 57361.8569 & 0.0006 & 0&0008 & 134 & 100 & 57366.3310 & 0.0006 & 0&0048 & 70 \\
36 & 57361.9292 & 0.0004 & 0&0044 & 174 & 101 & 57366.3987 & 0.0017 & 0&0038 & 34 \\
37 & 57361.9966 & 0.0003 & 0&0030 & 105 & 102 & 57366.4694 & 0.0005 & 0&0057 & 69 \\
41 & 57362.2698 & 0.0003 & 0&0011 & 67 & 114 & 57367.2879 & 0.0008 & $-$0&0011 & 53 \\
42 & 57362.3402 & 0.0003 & 0&0028 & 73 & 115 & 57367.3536 & 0.0009 & $-$0&0041 & 66 \\
43 & 57362.4099 & 0.0002 & 0&0037 & 107 & 116 & 57367.4225 & 0.0010 & $-$0&0040 & 51 \\
44 & 57362.4779 & 0.0003 & 0&0030 & 265 & 120 & 57367.7003 & 0.0009 & $-$0&0014 & 67 \\
45 & 57362.5484 & 0.0002 & 0&0047 & 435 & 121 & 57367.7692 & 0.0018 & $-$0&0012 & 68 \\
46 & 57362.6162 & 0.0002 & 0&0037 & 409 & 129 & 57368.3244 & 0.0005 & 0&0038 & 50 \\
47 & 57362.6888 & 0.0003 & 0&0075 & 94 & 130 & 57368.3928 & 0.0010 & 0&0035 & 52 \\
48 & 57362.7546 & 0.0005 & 0&0045 & 72 & 187 & 57372.2978 & 0.0008 & $-$0&0115 & 46 \\
49 & 57362.8227 & 0.0004 & 0&0038 & 73 & 188 & 57372.3535 & 0.0015 & $-$0&0246 & 46 \\
50 & 57362.8931 & 0.0005 & 0&0055 & 73 & 189 & 57372.4312 & 0.0013 & $-$0&0156 & 47 \\
51 & 57362.9640 & 0.0010 & 0&0076 & 24 & \multicolumn{1}{c}{--} & \multicolumn{1}{c}{--} & \multicolumn{1}{c}{--} & \multicolumn{2}{c}{--} & \multicolumn{1}{c}{--}\\
\hline
  \multicolumn{12}{l}{\commenta BJD$-$2400000.} \\
  \multicolumn{12}{l}{\commentb Against max $= 2457359.4490 + 0.068772 E$.} \\
  \multicolumn{12}{l}{\commentc Number of points used to determine the maximum.} \\
\end{tabular}
\end{center}
\end{table*}

\end{document}